\documentstyle[12pt]{article}

\begin{document}
     
\begin{flushright}
gr-qc/0012049 \\
Nature 408 (2000) 661-664
\end{flushright}

\begin{centering}
\bigskip
{\Large \bf Quantum theory's last challenge\footnote{This is the 
preprint version of an article that appeared in the issue 6813 (volume 408)
of Nature, as part of a 3-article celebration of the 100th anniversary
of Planck's solution of the black-body-radiation problem.
The text of the published version is not exactly the same; moreover,
the published version contains 4 illustrations, which are not included here.
The article is intended for a wide non-technical readership,
but it is not really self-contained, since it relies
on points discussed by Anton Zeilinger in another article composing
the celebration.}}\\
\bigskip
\bigskip
\bigskip
{\bf Giovanni AMELINO-CAMELIA}\\
\bigskip
Dipartimento di Fisica, Universit\`{a} ``La Sapienza", P.le Moro 2,
I-00185 Roma, Italy
\end{centering}
\vspace{1.4cm}
\begin{center}
{\bf ABSTRACT}
\end{center}

{\leftskip=0.6in \rightskip=0.6in

{\small
Quantum mechanics is now 100 years old and still going strong. 
Combining general relativity with quantum mechanics is the last 
hurdle to be overcome in the ``quantum revolution".
}
}

\newpage
\baselineskip 12pt plus .5pt minus .5pt
\pagenumbering{arabic}
\pagestyle{plain}

\baselineskip = 12pt

This year we celebrate 100 years of quantum theory, 
and in particular the anniversary of an announcement 
made by Max Planck at a meeting of the German Physical Society 
on 14 December 1900.  Planck was interested in the nature of 
radiation emitted by hot objects, and in 1900 he devised a theory 
that described all of the experimental evidence, but that required 
a radical new concept: energy is not emitted or absorbed continuously, 
but in discrete amounts, called quanta. 
At the time, Planck was not aware of the profound consequences 
of his work, but gradually physicists realized that they needed 
quantum concepts to understand the structure of all matter and radiation.

Over the past century there have been many successful tests of 
quantum mechanics (see accompanying commentary by Anton Zeilinger 
in this issue~\cite{zeilinger}). Experiments have confirmed even some 
of the most counterintuitive predictions of quantum theory, 
including ``particle-wave duality", the idea that a particle 
should be treated as a wave. But some physicists still wonder 
whether quantum theory is a truly fundamental ingredient of the 
laws of nature, or just a convenient description of  some aspects
 of the microscopic world. It is still possible that quantum mechanics
 is an approximation to a more fundamental theory, just as Newtonian 
gravity is a special case of the more accurate description of gravity
 and the relationship between space and time provided by Einstein's 
general relativity theory.

The biggest challenge to accepting quantum mechanics as a fundamental 
theory of nature is that despite 70 years of 
attempts~\cite{stachisto} it has 
still not been integrated with the classical theory of general relativity. 
 Most of the time, the subtleties of quantum mechanics can be safely ignored
 by general relativity: gravity drives the expansion of the Universe and the
 formation of galaxies, whereas quantum theory reigns supreme at the atomic 
scale. But there are times when quantum mechanics cannot be excluded.
 For example, a theory unifying gravity and quantum mechanics is 
required to understand the ``Big Bang" - the first few moments of 
the universe when gravitational interactions were very strong and 
the scales involved were all microscopic. 

Another area of concern is the lack of experimental evidence on the 
interplay between quantum theory and general relativity. 
The structure of the two theories does allow for  situations in
 which neither can be neglected, but it is extremely hard to create
 the required conditions in a laboratory. This is crucial if we are
 to test any of the theoretical ideas that propose to unify 
quantum mechanics and general relativity.

\bigskip
\noindent
{\bf The incomplete revolution.}

The ``relativity revolution" and the ``quantum revolution" are among the 
greatest successes of twentieth century physics, yet the theories they 
produced appear to be fundamentally incompatible. 
General relativity remains a purely classical theory: 
it describes the geometry of space and time as smooth 
and continuous whereas quantum mechanics divides everything 
into discrete chunks.  The predictions of the two theories 
have been confirmed in a large number of experiments, but 
each of these experiments is relevant for only one or the 
other of the two theories because of the different scales involved.  
Physicists' discomfort with this situation is perhaps best described 
by Rovelli's characterization~\cite{crincomplete} of the 1900s 
as ``the century of the incomplete revolution", suggesting 
there may be a greater revolution to come.

Part of the underlying theoretical incompatibility arises 
from the way the two theories treat the geometry of space 
and time. In quantum mechanics, space-time has the role of 
a fixed arena within which one describes the evolution of 
various ``quantum observables", such as the position of particles. 
But in general relativity, space and time are dynamical quantities
 that can respond to and influence other physical processes, such 
as the movement of planets. So, in principle, the particles in a 
quantum mechanics experiment affect the evolution of space-time 
through the gravitational fields generated by their energies. 
In practice, the particles are usually so small that they have
 a negligible effect on space-time dynamics, 
and general relativity can be ignored. 
This has been the case in all successful test of quantum mechanics.

\bigskip
\noindent
{\bf Approaches to quantum gravity.}

The search for a more fundamental theory, in which the 
incompatibility between  general relativity and quantum mechanics 
be resolved, can take two different paths. The most drastic option 
is to look for an alternative theory based on principles that are 
profoundly different from the principles of quantum mechanics. 
In such a theory quantum mechanics would only be meaningful as 
an approximation~\cite{polonpap,thoonew}.  
The other option is to maintain the existing principles of 
quantum mechanics but develop a more fundamental theory that 
can answer questions of a different type~\cite{crincomplete}, 
such as the ones emerging with a dynamical space-time. 
The situation in which quantum mechanics is abandoned at 
the level of the more fundamental theory is a much harder 
challenge, because one would need to rewrite the rulebook, 
and no serious candidate has yet emerged. The second option 
is more easily explored: at least one has quantum theory as 
a starting point. 

There are currently two mature theories based on quantum mechanics 
that attempt to unify general relativity and 
quantum mechanics: ``canonical quantum 
gravity"~\cite{crincomplete,clqg} 
and ``superstring theory"~\cite{string1,string2}. 
Although they share 
a common goal these two theories are quite different in the way 
they approach the technical and physical problems that emerge 
when building a quantum picture of gravity and space-time. 
In particular, they differ in the way they handle the mathematical 
infinities that naturally occur in quantum descriptions of 
gravitational fields. These infinities are not peculiar to 
quantum gravity: for example, they also appear in the 
quantum description of electromagnetic fields 
(the subject of quantum electrodynamics or QED) 
but in QED they can be removed by a technique known as 
perturbative renormalization. Thanks to this technique, 
QED has become one of the most accurate and best-tested 
theories of modern physics. 
But perturbative renormalization does not work for the 
quantization of Einstein's theory of gravity: 
the mathematical infinities are too persistent.  
In canonical quantum gravity ``non-perturbative renormalization" 
is being tried.  In ``superstring theory" perturbative renormalization
 can be made to work if it is assumed that the particles we perceive
 (with the relatively low resolution we presently have) as point-like
 are actually more properly described using one-dimensional entities 
(strings) that exist in a space-time with ten  dimensions. These ten
 dimensions include the four we ordinarily perceive (three of space 
and one of time) plus six other dimensions that would only show up 
in experiments at extremely small sales. None of these extra
 dimensions, or any of the string-like states, have yet been observed.

The two approaches also differ in the way they deal with the most 
fundamental challenge to the unification of general relativity and
 quantum mechanics: replacing the fixed space-time arena with 
a dynamical version. In ``superstring theory" the idea of a fixed 
space-time arena is conserved, but the theory includes some 
dynamical variables that also describe space-time. The end 
result is a dynamical space-time 
(which, in a sense,  is the ``sum" of the classical fiducial 
starting point and some dynamical quantum space-time effects), 
but some physicists feel that the starting point provided by 
the fiducial space-time arena does not fully embrace the lessons
 taught by general relativity.
 In this respect ``canonical quantum gravity" is 
more ambitious: space-time is fully dynamical from the start. 
Unfortunately it is not yet clear how this quantum dynamical 
space-time will describe the important limiting cases 
(corresponding to the majority of the viable experimental situations)
 where space-time does behave as a classical and fixed arena.

\bigskip
\noindent
{\bf A quantum theory of space-time.}

One of the most exciting possibilities emerging 
from ``canonical quantum gravity", ``superstring theory" 
and other (less developed) approaches to the 
unification of general relativity and quantum mechanics 
is the idea of a space-time that is itself quantized.  
This could involve a discretization of space-time and/or 
a quantum ``uncertainty principle" similar to that found 
in ordinary quantum mechanics. In a space-time with a 
quantum uncertainty principle it would not be possible
 to measure accurately the distance between two space-time points, 
just as in ordinary quantum mechanics it is not possible to measure
 (simultaneously) the position and momentum of a particle within 
the classical space-time arena.

The quantum description of space-time would require a profound 
renewal of fundamental physics. For example, it might have 
unexpected effects on the  propagation of particles. These effects 
would only emerge at the small scale of the Planck length (which is 
not easily explored experimentally).  
A useful analogy is the one of the surface of a wooden table. 
 We usually perceive the surface as perfectly flat, but if we 
look at the table with a microscope (which allows us to acquire
 sensitivity to its short-distance structure) it becomes clear 
that it is not exactly flat. The flatness we ordinarily perceive 
is some sort of averaging over the short-distance irregularities 
of the table's surface. If we take a ball with diameter much larger 
than the short-distance scale of roughness of the table and roll 
it over the surface, we find no evidence of the roughness, but if
 we use a ball whose diameter is not  much larger that the scale 
of roughness we will see disturbances in the path of the ball
 resulting from its interaction with the roughness. 

Similarly, a rich short-distance structure in space-time may 
have important effects on the propagation of particles, in a 
way that depends on the wavelengths of the particles. 
(Because of the wave-particle duality of quantum mechanics, 
all particles can be characterized by a wavelength, which is 
inversely related to their momentum.) In the tabletop analogy
 the diameter of the balls plays the role of the particle wavelength.
 For the particles we observe experimentally there should be very 
little wavelength dependence, but for particle wavelengths comparable
 with the tiny Planck length the phenomenon of particle propagation 
would be severely affected (imagine rolling a ball with diameter 
smaller than the scale of surface roughness over the table).

\bigskip
\noindent
{\bf Hints of a new revolution?}

A key ingredient in Planck's revolutionary breakthrough was 
provided by Ludwig Boltzmann's earlier work, particularly 
his interpretation of the second law of thermodynamics as a
 statistical law rather than an absolute law of nature. 
At the time no one could have imagined that Boltzmann's work
 would eventually play a role in studies proving the 
inconsistency of classical mechanics. Is it possible that we
 already have some of the ingredients of a new revolution today? 
In recent years, several theoretical ideas have emerged that
 would require a change of perspective as radical as Boltzmann's. 
These include the theories that quantize space-time discussed 
above, which explain the classical properties of space-time 
that are perceived by long-wavelength particles as a sort of 
average over the richer microscopic structure of quantum space-time.
 This is not unlike Boltzmann's description of the second law 
of thermodynamics as a statistical outcome of the properties of 
the many microscopic constituents of a thermodynamic system.

On the experimental side the crucial input for Planck's breakthrough 
came from puzzling data on the radiation emitted by hot bodies 
in thermal equilibrium with their surroundings. The light from 
these bodies is a mixture of different frequencies (colours). 
The classical formula linking the radiation emitted at different 
frequencies with the temperature of the hot body works well for 
radiation at low frequencies, but disagrees with experimental 
data at high frequencies. Planck observed that this puzzle could 
be solved by assuming that energy could be emitted or absorbed 
only in discrete amounts, and the quantum of energy was born.  

Among the experimental puzzles confronting fundamental physics 
today there are a few that could have the potential to lead to 
a profound renewal of fundamental physics.  A relevant illustrative
 example is provided by the observation of cosmic rays above the 
so-called Greisen-Zatsepin-Kuzmin (GZK) limit~\cite{gzk1,gzk2}.  
Cosmic rays are particles emitted by distant active galaxies. 
These rays produce showers of elementary particles when they 
pass through the Earth's atmosphere. Before reaching us ultra-high-energy 
cosmic rays travel gigantic (cosmological) distances and astronomers
 would expect interactions with photons in the cosmic microwave 
background (the faint glow left over from the big bang) to cool 
them down, producing a cut off at higher energies. The discovery 
of cosmic rays with energies above this GZK cut-off~\cite{gzkdata} 
apparently originating from outside our Galaxy, cannot be 
explained by current theories. 

To calculate the GZK limit theorists have to take into account 
Lorentz symmetry, a property of classical space-time . 
It has been suggested that the observation of cosmic rays above 
the GZK limit could be explained by assuming very small modifications
 to the predictions of Lorentz symmetry~\cite{colgla,kifu}. 
A plausible origin for such a modification comes from an 
unexpected source - the quantization of space-time.  
Remarkably, the value of the GZK limit predicted by calculations within 
some of the scenarios for quantization of space-time turns out to 
be consistent with the astronomical data~\cite{gactp2}. 

There are other plausible explanations for the cosmic-ray puzzle, 
but the explanation involving a quantum space-time is the only one
 that would simultaneously solve another similar 
puzzle~\cite{gactp2,nsblazars}.
 Astronomers have detected high-energy photons from distant astrophysical
 sources that, according to the classical picture of space-time, 
should not be able to reach us because they are expected to interact 
with the cosmic infrared background. This means that there should be 
a theoretical upper limit to the energies of the photons, similar to 
that for cosmic rays. Experimental evidence for such puzzles is just 
emerging, so it is too early to build a case for a quantum description 
of space-time.  Nonetheless their existence is intriguing, and the mere
 possibility of such effects promotes the study of  quantum space-times
 from the realm of science fiction to that of  proper science.

\bigskip
\noindent
{\bf The equivalence principle.}

Experiments that can observe the interplay between general relativity 
and quantum mechanics are not easy to design. The tiny scale of 
the Planck length at which these effects are expected to emerge, 
and the extreme weakness of gravity compared with the other forces 
of nature, have led many to believe that the study of quantum gravity
 would be beyond the reach of  doable experiments. 
Moreover, there appears to be a fundamental theoretical issue that 
affects the experimental study of gravitational forces at the quantum 
level.  The standard approach to the operative definition of forces in
 quantum mechanics is in conflict with the Equivalence Principle, one 
of  the cornerstones of general relativity.

In the study of all other forces, the operative definition of the
 fields involves experimental particles with two primary characteristics: 
their charge under the influence of the force field and their inertial mass.
 As shown in a celebrated paper by Bohr and Rosenfeld~\cite{rose},
the only way to extract accurate information on the force field is to 
use particles with large inertial mass and small charge. In the case of 
studies of gravitational fields this type of strategy would require 
particles with large inertial mass and small gravitational mass (which is 
the gravitational charge in the sense that it sets the strength of the 
gravitational forces felt by the particle), but the equivalence principle 
requires inertial and gravitational masses to be equal. 
Therefore the standard (Bohr-Rosenfeld) strategy appears to 
lead~\cite{stachisto,polonpap}
to a fundamental limitation on the accuracy with 
which gravitational fields can be measured 
(and operatively defined).

\bigskip
\noindent
{\bf The experimental frontier.}

Despite these many challenges, in the last quarter of the twentieth 
century physicists have started to make some progress in the experimental 
study of quantum gravity. A first significant step was taken in the 
mid-1970s with experiments that, while not being capable of testing 
quantum properties of space-time itself, studied how  the presence 
of strong gravitational fields affected the quantum mechanical 
behaviour of microscopic particles~\cite{cow,danature}.  
These experiments established that the gravitational fields 
generated by the Earth affect interference experiments in a 
way that is completely analogous to the influence of electromagnetic fields. 

The first experimental studies~\cite{ehns,koste,hupe}
with sensitivity to conjectured quantum properties
of space-time were proposed in the 1980s, 
when it was realized that 
experiments testing one of the fundamental symmetries of particle 
physics, known as CPT, were finally reaching sensitivity levels that
 could plausibly detect minute modifications of these symmetries
 caused by the quantum properties of space-time. The sensitivity 
of CPT experiments (in particular studies of particles known as 
neutral kaons~\cite{cplear}) 
has continued to improve over the past few
 decades.  So far they have provided no evidence of any quantum 
property of space-time. 

For several years tests of CPT symmetry were the only examples of
 experiments being analysed from the point of view of quantum 
space-time, but over the past few years more proposals have been 
put forward.  The most sensitive among these experiments are the
 mentioned searches for departures from Lorentz symmetry in 
cosmic-ray physics.  Besides the GZK limit, another key 
implication of Lorentz symmetry - the prediction that the
 time needed by a massless particle to propagate over a
 given distance should not depend on the wavelength of 
the particle - is going to be tested with extremely high 
accuracy by astrophysics experiments now in preparation. 
Quantum space-time might alter the propagation of gamma-rays 
collected from distant 
gamma-ray bursts~\cite{grbgac} - some of the
 most powerful explosions in the Universe. Within 5 or 10 years
 the next generation of gamma-ray telescopes, such as the GLAST
 space mission~\cite{glast}, 
will reach sensitivity levels sufficient
 for testing the type of minute wavelength-dependence which 
(as illustrated by the tabletop analogy described above)  
could be a property of quantum space-time.

In addition to tests of Lorentz and CPT symmetries, it might even 
be possible to explore directly the structure of space-time itself.
 Modern gravity-wave interferometers are built to detect tiny 
fluctuations in the distances between some test masses that 
might be caused by a passing gravity wave~\cite{saulson}. Gravity 
waves are ripples in the fabric of space-time predicted by 
Einstein's classical theory of 
gravity,
but they have never been detected. In a recent paper~\cite{gacgwi}, 
I observed that  gravity-wave interferometers could
also be used in attempts to detect
quantum fluctuations of space-time; in fact, any physical 
process that induces fluctuations in the distances between 
the test masses will affect the interference effect. 
Interferometers under construction, such as the LIGO 
detector in the United States and the VIRGO detector 
in Italy, could be sensitive~\cite{gacgwi} to fluctuations 
at scales that correspond to the Planck length, but 
this possibility depends very strongly~\cite{polonpap} on 
the actual mechanism of these fluctuations, and much
 theoretical work is still needed in order to establish 
which fluctuation mechanisms give rise to consistent 
pictures of space-time. 

\bigskip
\noindent
{\bf Prospects for the 21st century.}

After 100 years of quantum theory and more than 70 years of 
failed attempts to unify general relativity and quantum mechanics, 
we finally have some promising theoretical and experimental avenues 
for the investigation of the interplay between relativity and quantum theory.
 There is a strong theoretical desire to unite these two theories, 
but without a healthy experimental programme our best models would 
remain as mere speculation.  For now, the experiments are limited 
in number and their sensitivities are still only at levels that 
would correspond to rather optimistic estimates of the effects 
(estimates that rely on the assumption that the strength of some of 
the effects induced by the unification of general relativity and 
quantum mechanics will only be mildly suppressed by the smallness 
of the Planck length). Nonetheless, it seems reasonable to expect 
that a growing number of experimental and theoretical ideas will 
emerge over the 21st century and will eventually allow us to bring 
clarity into the challenge to quantum mechanics presented by the 
need to come to terms with general relativity. 

\bigskip
\section*{Acknowledgements}
The writeup of this article went through several rounds of editing.
I felt I should use this opportunity to convey to scientists from other
fields an exciting picture of frontier research within and beyond
quantum mechanics, but some aspects of the subject are very technical 
and I feared that the final result might be disappointing in all
respects: in spite of 
painful oversimplifications and painful compromizes on factual
accuracy I could have missed the target of a piece readable by
non-experts. In the end a draft that I could live with did mature, 
but only thanks to several rounds of feed-back and precious guidance 
from John Stachel and Sarah Tomlin. 
The generous contribution of John and Sarah is very greatfully
acknowledged.

\bigskip
\bigskip
\bigskip

\baselineskip 12pt plus .5pt minus .5pt

\baselineskip = 12pt
     
\vfil

\end{document}